\documentstyle[preprint,aps]{revtex}
\input epsf
\begin{document}
\draft
\title{Analysis of Charge Asymmetry in Rare Dilepton $B$ Decays}
\author{Dongsheng Liu}
\address{Department of Physics, University of Tasmania\\
Hobart, AUSTRALIA 7001}
\date{1 March, 1995}
\maketitle

\begin{abstract}
We analyze forward-backward charge asymmetry of the lepton production
in rare decays $B\rightarrow X_s l^+l^-$ and
$B\rightarrow K^* l^+l^-$, including vector-resonance effects.
Certain regions of phase space, in which the asymmetry
is sensitive to individual short-distance coefficients,
are pointed out. In particular, we suggest a method to test
the coupling of the leptonic axial vector current to
the left-handed quark current experimentally.
\end{abstract}


%\date{#1} for dating receipt at Editorial Offices of APS

\section{Introduction}
Rare $B$ decays have been the focus of many experimental
and theoretical considerations\cite{xsphoton,rare}.
This is due to the amount of information on the standard model
that can be extracted from these processes.
The rare decays proceed through flavor-changing neutral current
diagrams that are absent at the tree level and thus provide
a good probe of the standard model at the loop level.
For example, recent measurements of photon penguin
induced processes $B\rightarrow X_s \gamma$ and
$B\rightarrow K^* \gamma$ by the CLEO group\cite{xsphoton}
are consistent with the standard model predictions.
Likewise, the dilepton decay $B\rightarrow X_s l^+l^-$
and corresponding exclusive modes will provide tests of
the $Z$ boson penguin and $W$ boson box diagrams
which take over the photonic penguin diagram
for large top quark mass.
On the other hand, unlike the radiative $B$ decays,
the dileptonic process are strongly dependent on the top quark
mass $m_t$ through the dominant internal top quark line, so that
a comparison between theoretical calculations as a
function of $m_t$ and experimental data may lead to constraints on
the top quark mass. Such an exercise has been done with the current
upper limit of ${\rm Br}(b\rightarrow s \mu^+\mu^-)<5.0\times 10^{-5}$
($90$\% CL), which gives the bound of $\sim 390$~GeV on
the top quark mass \cite{DTP93}.
These rare $B$ decays are also sensitive to quark mixing angles
$V_{td}, V_{ts}$ and $ V_{tb}$, hence their measurements yield
valuable information on Cabibbo-Kobayashi-Maskawa(CKM) matrix elements
and consequently shed some light on CP violation in the standard model.
In this work we concentrate on the rare dileptonic decay
$b \rightarrow s l^+l^-$ ($l$ is either an electron or a muon.).

While extensive investigations have been carried out for
the invariant mass spectrum of dileptons,
the authors of Ref.\ \cite{AGM} pointed out that to fully extract
short-distance coefficients both in magnitude and in sign from
experiments, one has to consider the angular distribution such
as forward-backward charge asymmetry of the $l^+$ production in
the decay $b\rightarrow s l^+ l^-$, either.
In addition, this asymmetry can be used to test the chirality of
the $b\rightarrow s$ transition.
The process $B\rightarrow K^* l^+l^-$ has been considered for
the small recoil case where heavy quark symmetries apply~\cite{DD,DLiu}.
Recently, the Dalize distribution of the exclusive rare dilepton decays
has been derived in terms of various form factors~\cite{Zurich}
and the sensitivity of it to possible new physics has been investigated.
In particular, it is shown that measurement of the transverse
polarization of $K^*$ meson yields information on
the coupling of the leptonic axial vector current to
the left-handed quark current.
It is the aim of this work to provide a comprehensive analysis of
asymmetry for both inclusive process $B\rightarrow X_s l^+l^-$
and exclusive channel $B\rightarrow K^* l^+l^-$
in the context of the standard model.
We attempt to probe kinematic
regions in which asymmetries are sensitive to individual
short-distance coefficients. Thus measurements of these asymmetries
provide tests of short-distance physics which is sensitive to
extensions of the standard model. In this context it is suitable
to consider the {\sl integral} asymmetry as we
have flexibility to make various phase space cuts and read off
the corresponding `partial integrated' asymmetry.

The forward-backward asymmetry
results from the interference of leptons produced by
the vector current with that by the axial vector current (see Eq.~(10)).
This is because it is a parity violating effect.
In the $b\rightarrow s l^+ l^-$,
the axial vector
current is mainly due to the $Z$ boson penguin and $W$ boson box
diagrams in which top quarks are involved.
The vector current arises not only from these two diagrams along
with the photon penguin one, but also from
loops containing instead charm quarks.
Extensions of the standard model lead to extra diagrams.
In models with two Higgs doubles, for example,
there is a charged scalar particle. The
charged scalar coupling to quarks can substitute $W$ boson
in the $Z$ boson penguin diagram.
In addition to these, resonances such as the $J/\psi$ and $\psi'$,
created by the neutral four-quark operator $bsc\bar c$,
contribute to the vector current via the $V\rightarrow l^+ l^-$.
As shown later on, different sources of the vector current manifest
themselves in different regions of phase space, whereas the axial
vector current is constant overall. For the very low $q^2$
the photonic penguin dominates, while the $Z$ pengiun and $W$ box
becomes important towards high $q^2$. Furthermore, we shall
demonstrate in this work that the short-distance component of the
vector current suffers from cancellation dramatically in the middle
$q^2$ region. Actually, the pure top quarks effect vanishes
at an invariant mass close to the $J/\psi$ mass and
the analogy can be made with nonresonant contributions near the $\psi'$.
Thus
the asymmetry over there, roughly speaking, contains
to two factors; one is from top quarks responsible for
the axial vector current, and another is from charm quarks
for the vector current. The former is  short-distance and
sensitive to new physics, while the latter, we believe,
remains unchanged in extensions of the standard model.
As it is not renormalized under QCD, the effective axial vector
vertex can be calculated reliably.  On the other hand,
calculations show that the Wilson coefficient of the
$bsc\bar c$ operator, denoted by $a_2$,
is sensitive to the QCD scale parameter and the
renormalization point and scheme beyond
the leading logarithmic approximation~\cite{a1a2}.
But it is hopeful phenomenologically
to extract this coefficient quite precisely from nonleptonic
$B$ decays. With the top mass from the Fermilab
and the value of $a_2$ determined by the data of the
$B\rightarrow X_s J/\psi$,
the standard model prediction for forward-backward asymmetry
of the middle region is insensitive to
the dependence on the QCD renormalization parameters.
Meanwhile, measurements of the asymmetry offer us an
opportunity to detect the axial current in the
$b\rightarrow s l^+ l^-$ and to set constraints on
extensions of the standard model.
We suggest therefore that one should home in on
the resonant region for revealing the effect of
the axial vector current and short-distance physics in general.
This is contrast to the
decay rate spectrum where one has to `cut out' $J/\psi$ and $\psi'$
resonances to probe the short-distance effect.

This paper is organized as follows.
We will briefly review the effective Hamiltonian for
the flavor-changing neutral-current-induced transition
$b\rightarrow s l^+ l^-$ in the following section.
Then we proceed into the analysis of forward-backward asymmetry
of the inclusive process $B\rightarrow X_s l^+ l^-$ in Sec.\ III.
The exclusive channel $B\rightarrow K^* l^+ l^-$ is also examined
in the section, by employing heavy quark symmetry to determine
form factors from measurements of semileptonic $D$ decays.
Sec.\ IV is devoted to relevant discussions
along with our conclusion.  An Appendix contains some useful
integral formulae occurred in the main text.

\section{Effective Hamiltonian}

Let us begin with an effective Hamiltonian relevant to flavor-changing
one-loop processes
$b\rightarrow s l^+ l^-$  \cite{dilep},
\begin{equation}
H_{{\rm eff}}=\frac{G_F}{\sqrt{2}}\left( \frac \alpha {4\pi s_W^2}\right) [\bar
s%
\Gamma _\mu ^Ab\;\bar l\gamma ^\mu (1-\gamma _5)l+\bar s\Gamma_\mu ^Bb\;%
\bar l\gamma ^\mu (1+\gamma _5)l],
\end{equation}
with effective vertices,
$$
\Gamma_\mu ^{A(B)}=A(B)\gamma_\mu (1-\gamma _5)-im_bs_W^2F_2\sigma_{\mu
\nu }q^\nu (1+\gamma _5)/q^2.
$$
The coefficient functions are given by
\begin{equation}
A(B,F_2)=\displaystyle\sum_{q=u,c,t}U_q\;A_q(B_q,F_2^q),
\end{equation}
in which $U_q=V_{qs}^{*}V_{qb}$ is the product of the relevant
elements of the CKM matrix.
As long as the $u$ quark is ignored, one has $U_c+U_t=0$
from unitarity of the CKM matrix.
The heavy particles with masses much larger than the
physical scale $\mu\approx m_b$ are
integrated out and their masses are absorbed
into coefficient functions evaluated in two
steps~\cite{MW,QCD,misiak,munchen};
first one computes these functions at a renormalization point
about the heavy mass and then scales them down to the order of
the $b$ quark mass using the renormalization group equation.
In the standard model
the coefficients $A_t$ and $B_t$ at the scale equal to
the $W$ boson mass, for example, take the forms (with $x=m_t^2/M_W^2$)
\begin{equation}\begin{array}{rl}\displaystyle
A_t(x,1)-B_t(x,1)=&
\displaystyle\frac{1}{4}
\left[x+3\frac{x}{1-x}+3\frac{x^2}{(1-x)^2}\ln x\right],
\\ \\ \displaystyle
B_t(x,1)=&\!
-\left[\displaystyle
\frac{1}{2}x+\frac{3}{4}\frac{x}{1-x}
+\frac{3}{4}\frac{2x^2-x}{(1-x)^2}\ln x
-\frac{4}{9}(\ln x+1)\right.
\\ \\ & \quad\!\displaystyle\left.
+\frac{1}{36}\frac{82x^3\!-\!151x^2\!+\!63x}{(1-x)^3}
+\frac{1}{36}\frac{10x^4\!+\!59x^3-\!138x^2\!+\!63x}{(1-x)^4}\ln x
\right]s_W^2.
\end{array}\end{equation}
When the scale goes down, we notice first of all that
as the four-fermion operator with the axial vector current
of leptons has a vanishing anomalous dimension and does not
mix with other operators because of chirality,
the combination $A-B$ retains its value at the higher
mass scale and thus is independent of the choice of the lower scale $\mu$.
Secondly, at the level of the next-to-leading logarithm, we have
\begin{equation}\displaystyle
B_t(x,\eta)=B_t(x,1)+B_{{\rm QCD}}(\eta),
\end{equation}
in which~\cite{NLO}
$$
B_{{\rm QCD}}(\eta)=\displaystyle
-s_W^2\sum_{i=1}^6 \left[\left( \frac{\pi}{\alpha_s(M_W)}p_i + s_i \right)
					    (1-\eta^{-a_i-1}) +r_i\; (1-\eta^{-a_i})
\right ]
$$
enters through the mixture of the current-current and
the QCD penguin four-quark operators. Here the scale-dependent
parameter
$\eta=\displaystyle\frac{\alpha_s(\mu)}{\alpha_s(M_W)}$.
Finally, the magnetic-moment operator has nonzero anomalous dimension and
mixes with other operators; the coefficient for it is given by
\cite{misiak}
\begin{equation}\begin{array}{rl}\displaystyle
F^t_2(x,\eta)=&\displaystyle
\eta^{-16/23}\left[
\frac{1}{12}\frac{8x^3+5x^2-7x}{(1-x)^3}
+\frac{1}{2}\frac{3x^3-2x^2}{(1-x)^4}\ln x\right.
\\  & \displaystyle\left.\quad\quad\quad\quad
+(\eta^{2/23}-1)
\left(\frac{1}{3}\frac{x^3+5x^2-2x}{(1-x)^3}-\frac{2x^2}{(1-x)^4}\ln x
\right)\right]
\\ & \displaystyle
+2\sum_{i=1}^8 h_i\; \eta^{-a_i} .
\end{array}\end{equation}
With $\eta=1.75$ and $m_t=180$~GeV obtained by averaging the data of
CDF and D0 groups\cite{CDF}, we have the values of the coefficient
functions as follows
\begin{equation}
A_t=2.09, \quad B_t=-0.126, \quad s^2_W F_2^t=-0.153.
\end{equation}
Recent measurements of $B\rightarrow X_s \gamma$ by the CLEO group
\cite{xsphoton} result in a constraint on the magnitude of $F_2^t$,
which is close to the value of the standard model.
Likewise, the dilepton decay $B\rightarrow X_s l^+l^-$
provides a unique test of the coefficients $A$ and $B$
which are absent in the radiative process.

In addition to the top quark, the decay $B \rightarrow X_s l^+l^-$
involves the charm quark and as well known,
its total rate is actually dominated by the $c\bar c$ resonances.
The contribution of these resonances such as $J/\psi$ and $\psi'$
to the effective Hamiltonian
Eq.\ (1) may be taken into account by the usual form of
vector-meson dominance \cite{DTP,long,Paver}
$$\displaystyle
H_V=\frac{G_F}{\sqrt{2}} U_c Q_c\left(\frac{ef_V}{M_V}\right )^2
\frac{a_2}{q^2-M_V^2+iM_V\Gamma_V}\bar s\gamma_\mu (1-\gamma _5)b
\;\bar l\gamma^\mu l,
$$
where $M_V$ is the mass of the vector intermediate state and
$\Gamma_V$ its full width.
The decay constant is defined such that
$f_V\epsilon_\mu=\langle 0|\bar c\gamma_\mu c|V(\epsilon)\rangle$ and will be
determined by the measured partial width for decays to lepton pairs
\cite{pdg}
$$\Gamma(V\rightarrow l^+l^-)
=\displaystyle\frac{4\pi}{3}\frac{(Q_c\alpha)^2}{M_V^3}f_V^2,$$
with $Q_c=2/3$.
In this work we treat the coupling for the neutral $b\bar s c\bar c$ four-quark
operator $a_2=c_1 + c_2/3$ as a phenomenological parameter and use
the CLEO data $\mid a_2 \mid =0.26\pm 0.03$
\cite{a2}, which is close to the value determined from
a fit to the rate for the semi-inclusive process
$B \rightarrow X_s J/\psi$\cite{DT}. The choice of a negative $a_2$
is in consistence with Breit-Wigner phase $\varphi=0$
\cite{phase}. However, consequences of assuming a different phase
for $a_2$ can be derived  without any difficulty
in principle.
Hence we can read off the vector resonance component
for Eq.\ (2),
\begin{equation}\displaystyle
A_V=B_V=\displaystyle
\frac{16\pi^2}{3} \left(\frac{f_V}{M_V}\right )^2
\frac{a_2s_W^2}{q^2-M_V^2+iM_V\Gamma_V}.
\end{equation}

Moreover, there is an effect arising from quarks active at the $\mu$ scale,
which contribute via penguin diagrams with insertion of four-quark
operators. It is represented by
\begin{equation}\begin{array}{rl}
A_f=B_f=s_W^2 &\displaystyle
\left[(a_2+a_4+a_6)\; \phi(m_c^2/m_b^2,\; q^2/m_b^2)
	-\frac{a_3+a_4+a_6}{2}\; \phi(1,\; q^2/m_b^2)\right.
\\  &\displaystyle \left.
	-\frac{a_3}{2}\; \phi(0,\; q^2/m_b^2)+\frac{2}{3} (a_4+a_6) \right],
\end{array}\end{equation}
where $a_3=c_3/3+c_4,\; a_4=c_3+c_4/3,\; a_6=c_5+c_6/3,$ and
the function $\phi$ comes from the one-loop matrix elements of
the four-quark operators and is given by
$$\phi(r_q,s)=\!\left\{\!\begin{array}{l}\displaystyle
 \frac{4}{3}\ln r_q\!-\frac{8}{9}\!-\frac{4}{3}\frac{4r_q}{s}\!
+\!\frac{2}{3}\sqrt{1\!-\!\frac{4r_q}{s}}\left(\!2\!+\!\frac{4r_q}{s}\right)\!\!
 \left(\!\ln\frac{1\!+\!\sqrt{1\!-\!4r_q/s}}{1\!-\!\sqrt{1\!-\!4r_q/s}}
	  \!-\!i\pi\!\right)
 \quad\!\! (\frac{4r_q}{s}<1);
 \\ \\ \displaystyle
 \frac{4}{3}\ln r_q-\frac{8}{9}-\frac{4}{3}\frac{4r_q}{s}
+\frac{4}{3}\sqrt{\frac{4r_q}{s}-1}\left(2+\frac{4r_q}{s}\right)
 \arctan\frac{1}{\sqrt{4r_q/s-1}}
 \quad (\frac{4r_q}{s}>1).
\end{array}\right.
$$
Numerically, we have $a_3=0.0211,\; a_4=-0.0024,$ and $a_6=0.0028$
for $\eta=1.75.$ Comparing them with $a_2$ shows that
the charm quark practically dominates the effect of these active quarks.
Since the coupling of them to $l^+l^-$ is of vector-type,
the combination $A-B$ remains unchanged after they are included.
We remind the reader that the short-distance coefficients listed above
are expressions in the NDR scheme.
The scheme independence of the sum of these coefficients has been
explicitly demonstrated by the authors of Ref.~\cite{misiak}.

\section{Forward-backward asymmetry in dilepton $B$ decays}
\subsection{$B \rightarrow X_s l^+l^-$}

The differential forward-backward asymmetry of the $l^+$ production
that we are to compute is defined by
$$d\Gamma_{{\rm FB}}(q^2)=
  \int^1_0\; d\Gamma(\cos\theta_l)-\int^0_{-1}\; d\Gamma(\cos\theta_l),
$$
in which $\pi-\theta_l$ is the polar angle of the $l^+$
with respect to the direction of motion of the decaying meson
in the $l^+l^-$ frame. For the inclusive process
$B \rightarrow X_s l^+l^-$ induced by the Hamiltonian in Eq.\ (1),
it takes the form
\begin{equation}\displaystyle
\frac{d\Gamma_{{\rm FB}}}{dq^2}
= \displaystyle\frac{1}{4 \pi^3}
  \left ( \frac{G_F}{\sqrt{2}} \right )^2
  \left( \frac \alpha {4\pi s_W^2}\right)^2
  \frac{\lambda^2\;q^2}{m_b}
\left ( \left | B + \frac{m_b^2}{q^2} s^2_W F_2 \right |^2
	  -\left | A + \frac{m_b^2}{q^2} s^2_W F_2 \right |^2
\right ),
\end{equation}
where
$\lambda
=\displaystyle\frac{1}{2m_b}\sqrt{(m_b^2+m_s^2-q^2)^2-4m_b^2m_s^2}$,
with $m_b$ and $m_s$ the masses of the $b$ and $s$ quarks respectively.
The contribution of pure photonic penguin diagrams,
$\left |\displaystyle\frac{m_b^2}{q^2} s^2_W F_2 \right |^2$, which governs
the rate of radiative processes
$B \rightarrow X_s\;\gamma$, cancels out. Thus the measurement
of the asymmetry provides entirely independent tests of the standard model.
In Eq.\ (9) we have set the limit of $m_l=0$ because there is
no singularity at $q^2=0$; so
there is no mixture between left- and right-handed leptons and
the difference between electron and muon channels is negligible.
As noticed by the authors of Ref.\ \cite{AMM},
the forward-backward charge asymmetry is proportional to the coefficient
combination $(A-B)$ and may be large for
the very heavy top quark. Recalling that $(A-B)$ receives neither
the QCD renormalization nor the $c\bar c$ effect, we anticipate
that the asymmetry will become an effective measure for testing
the high energy scale physics.
To study its properties in detail, let us present a factorized form
for the asymmetry ({\it integrated} over the interval of $[0,\;q^2]$)
\begin{equation}\displaystyle
\Gamma_{{\rm FB}}(q^2)=
(A-B)\left( \frac \alpha {4\pi s_W^2}\right)^2
\left( I_{t}(q^2) + s_W^2I_{f}(q^2)
+ a_2s_W^2\displaystyle\sum_V I_V(q^2)  \right).
\end{equation}
Here $I_t$ stands for the contributions
of loops containing the virtual top quark along with QCD corrections,
for which we find
\begin{equation}\begin{array}{rl}\displaystyle
I_t(q^2)=
 \frac{m_b^5\mid U_t\mid^2 }{16\pi^3}
 \left( \frac{G_F}{\sqrt{2}} \right)^2
  &\left\{ \displaystyle
	  [A_t(x, \eta)+B_t(x, \eta)]
  \left[-\frac{1}{2}(1-r_s)^2s^2+\frac{2}{3}(1+r_s)s^3-\frac{1}{4}s^4\right]
\right.\\ &\left.\displaystyle
	 -2 s_W^2 F_2^t(x, \eta)
  \left[(1-r_s)^2 s-(1+r_s)s^2+\frac{1}{3}s^3 \right]
\right\},
\end{array}\end{equation}
with $s=q^2/m_b^2$ and $r_s=m_s^2/m_b^2$.  In Fig.\ 1 we plot
the integrated asymmetries arising from the photonic penguin
and the $Z$ penguin plus $W$ box diagrams.
They are rescaled in terms of the inclusive semileptonic
decay width of the $B$ meson
$$\displaystyle
\Gamma_{\rm s.l.}
=\displaystyle\frac{m_b^5\!\mid\! V_{cb}\!\mid^2 \!}{96\pi^3}
\left( \frac{G_F}{\sqrt{2}} \right)^2 \tilde f_{bc},
$$
with $\tilde f_{bc}$ numerically equal to $0.39$, in which we have
included the one-loop
QCD correction to the semileptonic $B$ decay (all plots in this paper
are rescaled in this way).
For low $s$, photonic penguin diagrams dominate $I_t(q^2)$, and
an opposite sign of $F_2^t$ to $A_t-B_t$ in the standard model
implies a positive asymmetry.
As $s$ increases, contributions of the $Z$ penguin and $W$ box diagrams
emerge and cancel with that of photonic penguin diagrams.
Consequently, the integrated asymmetry arising purely from
virtual top quarks turns out to be zero at
$$
s_0=1-y_0-\displaystyle\frac{1-4s_t}{9}
		\left[1+\displaystyle\frac{1}{y_0}
			   \left(1-\displaystyle\frac{1-4s_t}{9}
\right)\right]+{\cal O}(\frac{m_s^2}{m_b^2}),
$$
with
$$
y_0=\displaystyle\frac{1}{3}
\left[\displaystyle\frac{9}{2}(1-4s_t)
	+\displaystyle\frac{1}{2}(1-4s_t)^2
	+\displaystyle\frac{1}{27}(1-4s_t)^3
	+2(1-4s_t)\sqrt{(2-s_t)^2+2}
\right]^{1/3},
$$
and $s_t=-\displaystyle\frac{2s_W^2F_2^t}{A_t+B_t}$.
For $m_t=180\pm 12$~GeV and $\eta=1.75$ we have
$s_0=0.365\mp 0.010$ which, by chance, is slightly below
the $J/\psi$ peak point ($\approx 0.400$). The range of
this zero point due to the error bar of the $m_t$ is expected
to be small, considering that $(A_t+B_t)$ and $F_2^t$ both
depend rather weakly on the top mass.
We find that this zero point varies from $0.356\mp 0.008$
to $0.381\mp 0.012$ for the above top quark masses
when $\eta$ ranges from $1.50$ to $2.00$,
which incorporates uncertainties due to QCD scale parameter
as well as the renormalization subtraction point.
For the top component of the short-distance coefficient,
the difference between NDR and HV schemes amounts to
just about 5\%~\cite{munchen}. Thus we will see a slightly different
value of the $s_0$ in the HV scheme.
Toward the high $s$ region the $Z$ penguin along with $W$ box diagram
takes over in magnitude and we are left with
a {\em surplus} of the number of $l^-$ scattered in
the {\em forward} hemisphere in the $l^+l^-$ rest frame.

Other components of the pure short-distance effect,
coming from active quarks, are represented by
\begin{equation}\begin{array}{rl}
I_f(q^2)=\displaystyle
 \frac{m_b^5 \mid\! U_t\!\mid^2}{6\pi^3}
 \left( \frac{G_F}{\sqrt{2}} \right)^2
&\displaystyle\left\{
(a_2+a_4+a_6)\; \Phi(m_c^2/m_b^2,s)
	-\frac{a_3+a_4+a_6}{2}\; \Phi(1, s)
\right. \\ & \displaystyle\left.
    -\frac{a_3}{2}\; \Phi(0,s)
    +\frac{a_4+a_6}{2} \left[
    \frac{1}{2}(1-r_s)^2 s^2 -\frac{2}{3}(1+r_s) s^3 + \frac{1}{4} s^4
					\right]\right\},
\end{array}\end{equation}
with
\begin{equation}\begin{array}{rl}
\Phi(r_q,s)& = \displaystyle
 \left(\ln r_q-\frac{2}{3}\right)
 \left[ \frac{1}{2}(1-r_s)^2 s^2 -\frac{2}{3}(1+r_s) s^3 + \frac{1}{4} s^4
 \right]
 \\ & \displaystyle
-4r_q[(1-r_s)^2 s -(1+r_s) s^2 + \frac{1}{3} s^3 ]
 \\ & \displaystyle
\begin{array}{rl}
+16r_q^2 & [\;32r_q^2I_5(z)-16r_q(1+r_s+r_q)I_4(z) \\
	    &  + 2((1-r_s)^2+4r_q(1+r_s)-8r_q^2)I_3(z) \\
	    &  -((1-r_s)^2-8r_q(1+r_s))I_2(z)-(1-r_s)^2 I_1(z)],
\end{array}
\end{array}\end{equation}
where the explicit form for $I_k(z)$ is given in the Appendix
and $z=\sqrt{1-\displaystyle\frac{4r_q}{s}}$.
Figure 2 summarizes asymmetries coming from the short-distance
effects alone. The nonresonant $c\bar c$ component dominating active
quark effects becomes
substantial above the threshold $2m_c$. Unfortunately, the total
short-distance contribution suffers from a cancellation
between the top  and $c\bar c$ components,
which reduces strongly the surplus of $l^-$ of high $q^2$ in the forward
hemisphere. With the same reasoning, we expect a vanishing net result of
the short-distance effect as a whole somewhere in the middle region
of the phase space. This sort of zero points (denoted by $\bar s_0$)
is identical in the NDR and HV schemes.
To be precise, we evaluate the position of this point
for $m_t=180\pm 12$~GeV and $\eta=1.75$. It turns out to be
$\bar s_0=0.511\mp 0.025$, which comes closer to the $\psi'$ resonance
than the $J/\psi$.
We find a weak $\eta$ dependence of this zero point
and obtain a derivation less than 4\% for $\eta$ varying from
1.50 to 2.00 once again.

Now let us consider resonant effects arising from
vector intermediate states. We have
\begin{equation}\begin{array}{rl}\displaystyle
I_V(q^2)
=\frac{2 m_bf_V^2 \mid\! U_t\!\mid^2}{3\pi r_V}
\left( \frac{G_F}{\sqrt{2}} \right)^2\!\!
 &\displaystyle\!\! \left[
 [(1+r_s-r_V)^2-4r_s]s
-\frac{1}{2}[2(1+r_s)-r_V]s^2+\frac{1}{3}s^3
\right. \\ &\displaystyle\left.
+r_V\left[(1-r_V)^2+2(1+r_V)r_s-r_s^2\right]
\ln\sqrt{(1-\displaystyle\frac{s}{r_V})^2+\sigma_V^2}
\right. \\ &\displaystyle\left.
+{\cal O}(\sigma_V) \right],
\end{array}\end{equation}
with
$r_V=M_V^2/m_b^2$ and $\sigma_V=\Gamma_V/M_V$.
The resonant component $I_V(q^2)$ reaches its
maximum value at $s=r_V$. Because of the dependence on the real part
of the resonant propagator, as $s$ passes through $r_V$,
the differential asymmetry changes sign. There is , therefore,
a partial cancellation of the $I_V(q^2)$ and instead of a plateau
we will find a decreasing integrated asymmetry for $s$
across the mass of the vector resonance.

The integrated asymmetry of the $l^+$ production in the inclusive process
$B\rightarrow X_s l^+l^-$ is plotted in Fig.\ 3 and includes
the components of $J/\psi$ and $\psi'$ resonances.
It is characterized by prominent peaks at $q^2=M^2_{J/\psi}$ and
$M^2_{\psi'}$ respectively in the resonant region, as expected.
The contribution of the $\psi'$ resonance is around one third
that of the $J/\psi$.
In the $J/\psi$ resonant region, the top quark component
of the asymmetry is actually small, as explained previously.
It means that the product $a_2(A-B)$ dominates,
as far as the coefficients are concerned.
Unlike the decay rate, the $(A-B)$ is involved here linearly.
Alternatively stated, the short-distance effects as a whole
disappears at $\bar s_0$, which is near the $\psi'$.
The asymmetry, therefore, appears
in the shape of the real part of the resonant propagator.
Experimentally, $a_2$ has been reasonably well extracted from the CLEO data
and hopefully more precise data will come from the planned $B$
factories. As a result, the prediction for asymmetries near
these zero points (or resonances) can be made in a way insensitive to
the QCD renormalization parameter $\eta$.
Numerically, as $\eta$ ranges from $1.50$ to $2.00$,
$\Gamma_{\rm FB}(M^2_{J/\psi})$, for example, varies merely by $6.1\%$.
In our opinion, measurements of these asymmetries will be
an effective way to unambiguously test the combination $A-B$
arising purely from short-distance physics.
Meanwhile, as $A-B$ depends strongly on the top mass in
the standard model, it may be an independent way
to confirm the value of $m_t$ determined in different ways.

\subsection{$B \rightarrow K^* l^+l^-$}

We now turn our attention to the exclusive channel
$B\rightarrow K^*\;l^+l^-$. The effective quark current
$\bar s\Gamma _\mu ^{A(B)}b$ in question has two different structures;
the parametrization for the matrix element of $V-A$ currents in terms of
invariant form factors is
\begin{equation}\begin{array}{rl}
\langle p, \phi|\bar s\gamma ^\mu (1-\gamma _5)b |P\rangle = &
[a_{+}(q^2)(P+p)_\mu +a_{-}(q^2)(P-p)_\mu ]P^\nu \phi_\nu ^{*} \\
&+f(q^2)\phi_\mu ^{*}
 +ig(q^2)\epsilon _{\mu\nu\alpha\beta}\phi^{*\nu }P^\alpha p^\beta.
\end{array}\end{equation}
In analogy to this we have for the magnetic-moment operator
\begin{equation}\begin{array}{rl}
-\displaystyle\frac{i}{q^2}
\langle p, \phi|\bar s\sigma_{\mu \nu }q^\nu (1+\gamma _5)b |P\rangle = &
[\tilde a_{+}(q^2)(P+p)_\mu +\tilde a_{-}(q^2)(P-p)_\mu ]P^\nu \phi_\nu ^{*} \\
& +\tilde f(q^2)\phi_\mu ^{*}
  +i\tilde g(q^2)\epsilon _{\mu\nu\alpha\beta}\phi^{*\nu }P^\alpha p^\beta,
\end{array}\end{equation}
along with a condition of current conservation,
$$(M^2-m^2)\tilde a_+ + q^2 \tilde a_- + \tilde f =0.$$
Here $M\;(P_\mu)$ and $m\;(p_\mu)$ are masses (momenta)
of $B$ and $K^*$ mesons, respectively,
$\phi_\mu$ the polarization vector of the $K^*$ meson (satisfying
$\phi_\mu p^\mu =0$) and $q=P-p$ the momentum transfer into the dilepton.

The differential forward-backward asymmetry of
$B\rightarrow K^*\;l^+l^-$ reads \cite{DLiu}
\begin{equation}\displaystyle
\frac{d\Gamma_{\rm FB}}{dq^2}
= \displaystyle\frac{1}{64 \pi^3}
  \left ( \frac{G_F}{\sqrt{2}} \right )^2
  \left( \frac \alpha {4\pi s_W^2}\right)^2
  \frac{\lambda\;q^2}{M^2}
[ (\mid H_+^L\mid^2 -\mid H_-^L\mid^2 )
- (\mid H_+^R\mid^2 -\mid H_-^R\mid^2)],
\end{equation}
where necessarily now
$\lambda=\displaystyle\frac{1}{2M}\sqrt{(M^2+m^2-q^2)^2-4M^2m^2}$.
The helicity amplitudes appearing here are defined as
\begin{equation}
H_\pm^{L(R)}=A(B) h_\pm + m_bs_W^2F_2^t\tilde h_\pm,
\end{equation}
where $h_\pm\equiv f\pm\lambda M g$  \cite{KS}
and $\tilde h_\pm\equiv\tilde f\pm\lambda M \tilde g$.
Assuming the $B$ meson contains an on-shell $b$ quark of velocity $v$,
one finds
\begin{equation}
\tilde h_\pm=\frac{1}{q^2}(M-v\cdot p\mp \lambda)\; h_\pm.
\end{equation}
(see Ref.\ \cite{DD,DLiu} for detailed discussion.) With these relations we
may rewrite the differential asymmetry of Eq.\ (16) as
\begin{equation}\begin{array}{rl}\displaystyle
\frac{d\Gamma_{\rm FB}}{dq^2}&
= \displaystyle\frac{1}{64 \pi^3}
  \left ( \frac{G_F}{\sqrt{2}} \right )^2
  \left( \frac \alpha {4\pi s_W^2}\right)^2
  \frac{\lambda\;q^2}{M^2}(A-B)
\\ &\times\!\displaystyle
  \left[\mid\! h_+\!\mid^2\!\!
	   \left(\!A+\!B^*\!+2s_W^2F_2\frac{M\!-v\cdot p-\!\lambda}{q^2/m_b}\right)
	  -\mid\! h_-\!\mid^2\!\!
	   \left(\!A+\!B^*\!+2s_W^2F_2\frac{M\!-v\cdot p+\!\lambda}{q^2/m_b}\right)
\right].
\end{array}\end{equation}

Certainly, further analysis requires the knowledge of hadronic form
factors, but it is hard to calculate them directly from the first
principles of QCD.
In Ref.\ \cite{DD,DLiu} we have used heavy flavor symmetry
to relate the form factors for $B\rightarrow K^* l^+l^-$ decays with
those for $D\rightarrow K^*\; l^+ \nu$. In the leading order of
 heavy quark effective theory, form factors scale as \cite{IW42}
\begin{equation}\begin{array}{c}
\displaystyle
 f=\left [\frac{\alpha_s(m_b)}{\alpha_s(m_c)}\right ]^{-6/25}
  \sqrt{\frac{M}{M_D}}\;f_D, \\ \\ \displaystyle
 g=\left [\frac{\alpha_s(m_b)}{\alpha_s(m_c)}\right ]^{-6/25}
  \sqrt{\frac{M_D}{M}}\;g_D.
\end{array}\end{equation}
In this equation form factors are evaluated at the same
$v\cdot p$, namely that $f$ and $g$ at $q^2=M^2+m^2-2M v\cdot p$
are related to $f_D$ and $g_D$ at $q_D^2=M_D^2+m^2-2M_D v\cdot p$.
Thus the whole physical range of $q_D^2=[0, (M_D-m)^2]$
in $D\rightarrow K^*$  covers $q^2=[q_0^2, (M-m)^2]$
in $B\rightarrow K^*$, where $q_0^2=(4.07$~GeV$)^2$.
With the data available for $D\rightarrow K^*$
we are led to a reliable estimate of the asymmetry
in the region above the $\psi'$ resonance, on the basis of
momentum-to-momentum correspondence of form factors\cite{DLiu}.
However, extrapolating the data of $D$ decays to lower $q$
definitely needs the knowledge of the $q^2$-dependence for form factors.
In this work we will employ the pole dominance approximation,
\begin{equation}
 \left ( 1-\displaystyle\frac{q^2}{M'^2}\right )^n\; h(q^2)
=\left ( 1-\displaystyle\frac{q^2_{0}}{M'^2}\right )^n\; h(q^2_{0});
\quad\quad (h=f,g).
\end{equation}
Here $M'$ is the pole mass of the current involving
$b$ and $s$ quarks and we make no distinction between the vector
and axial vector masses.
While this form seems reasonable for the small recoil region of
$B\rightarrow K^*$ decays, as the end-point is close to the pole,
we hope it helps to figure out the properties of forward-backward
asymmetries in the resonant region or below.
Regarding poles of degree $n$, we make use of two schemes;
one assumes the monopole form for both $f$ and $g$
like the BSW model\cite{BSW}, and another follows the suggestion
of K\" orner and Schuler \cite{KS88}, namely $n=1$ for $f$,
but a dipole form of $n=2$ for $g$ (The latter is referred to as the dipole
scheme in this paper despite $f$ actually being of monopole form.).
Meanwhile, Eq.\ (20) will be used to obtain form factors at the
kinematical point $q_{0}$;
so with form factors $A_1(0)$ and $V(0)$
(they are linear combinations of $f_D$ and $g_D$)
of $D\rightarrow K^*(892)\; l^+ \nu$ measured
by E691, E687 and CLEO groups \cite{DK1}, one ends up with
$f_D(0)=-1.43$~GeV and $g_D(0)=0.695$~GeV$^{-1}$,
which translate to $f(q_0^2)=-2.64$~GeV
and $g(q_0^2)=0.455$~GeV$^{-1}$.

Before doing practical calculations, we perform
a simple analysis of integral asymmetries,
by using the monopole form factor along with
the limit of $M=M'$ and $m^2/M^2\ll 1$.
It ought to be appropriate for the region not close to the pole,
say around the $J/\psi$ mass or below.
In this case, the counterpart of Eq.\ (11) in the $K^*$ mode,
with necessarily $s=q^2/M^2$ for this moment,
has the form
\begin{equation}\begin{array}{rl}\displaystyle
I_t^{(K^*)}(q^2)=&\displaystyle
\!\frac{M^5\!\mid\! U_t\!\mid^2 \!}{128\pi^3}
 \left( \frac{G_F}{\sqrt{2}} \right)^2
 \left\{ f_0g_0 [A_t(x, \eta)+B_t(x, \eta)] s^2 \right.
 \\ &\left.\displaystyle
 -2s_W^2F_2^t(x, \eta)
  \left[ \left(\frac{f_0}{M}-\frac{Mg_0}{2}\right)^2\! s
	   -\frac{Mg_0}{2}\left(\frac{f_0}{M}+\frac{Mg_0}{2}\right)s^2
	   +\frac{1}{3}\left(\frac{Mg_0}{2}\right)^2\! s^3
\right] \right\}.
\end{array}\end{equation}
We have defined that $f_0=f(0)$ and $g_0=g(0)$.
Once again, competition between photon penguin and $Z$ penguin
plus $W$ box diagrams causes a zero point of $I_t^{(K^*)}$ at
\begin{equation}\displaystyle
\displaystyle\frac{3}{2}
\left(1-2\frac{(2-s_t)}{s_t}\frac{f_0}{M^2g_0}\right)
\left[
 1-\sqrt{
	   1-\frac{4}{3}
		\left(\frac{1-2\displaystyle\frac{f_0}{M^2g_0}}
				 {1-2\displaystyle\frac{(2-s_t)}{s_t}
							  \frac{f_0}{M^2g_0}
				 }\right)^2
	   }\;\right].
\end{equation}
Clearly, as $s_t < 1$ in the standard model, the condition
that $f_0$ is of opposite sign to $g_0$ guarantees
a real value of this root. This condition is satisfied physically because
the helicity relation $\mid h_+\mid<\mid h_-\mid$ has been well
confirmed in semileptonic decays\cite{KS}, namely that
the $V-A$ structure of the quark current is manifested
at the hadron level.
Numerically, with $f(q^2_0)$ and $g(q^2_0)$ given earlier
the zero point turns out to be $0.345\mp 0.07$ for
$m_t=180\pm 12$~GeV and $\eta=1.75$. It coincides with
the $J/\psi$ peak $M_{J/\psi}^2/M^2=0.344$.

Meanwhile, the correspondence to Eq.\ (12) reads
\begin{equation}\begin{array}{rl}
I_f^{(K^*)}(q^2)=\displaystyle
 \frac{M^5 \mid\! U_t\!\mid^2}{24\pi^3}
 \left( \frac{G_F}{\sqrt{2}} \right)^2 f_0g_0 \!
&\displaystyle\left[
(a_2+a_4+a_6)\; \Phi^{(K^*)}(m_c^2/M^2,s)
	- \frac{a_3}{2}\; \Phi^{(K^*)}(0,s)
\right. \\ & \displaystyle\left.
 -\frac{a_3+a_4+a_6}{2}\; \Phi^{(K^*)}(1, s) -\frac{a_4+a_6}{4} s^2
					\right],
\end{array}\end{equation}
with
\begin{equation}\displaystyle
\Phi^{(K^*)}(r_q,s) =
4r_qs-\frac{1}{2}(\ln r_q-\frac{2}{3})s^2
	 -16r_q^2[2I_3(z)-I_2(z)-I_1(z)].
\end{equation}
In analogy to the inclusive process, adding the active quark effect
to the top one ends up with a cancellation of the asymmetry once again.
We find a complete cancellation at $\bar s_0=0.451\mp 0.012$ for the same top
mass and $\eta$ as before.
In addition we work out the vector resonant contribution
which is
\begin{equation}\displaystyle
I_V^{(K^*)}(q^2)
=-\frac{Mf_V^2 \mid\! U_t\!\mid^2}{6\pi}
\left( \frac{G_F}{\sqrt{2}} \right)^2  f_0g_0
\left(
 \frac{s}{r_V}+\ln\sqrt{(1-\displaystyle\frac{s}{r_V})^2+\sigma_V^2}
+{\cal O}(\sigma_V)
\right),
\end{equation}
where now $r_V=M_V^2/M^2$.
We should remark that the above expressions illustrate
the features of the asymmetries in $B\rightarrow K^* l^+l^-$ ,
viz., the peaks in the resonant contribution,
the cancellation between $Z$ penguin and plus $W$ box diagrams,
and the plateau in the nonresonant $c\bar c$ effect above
the threshold $2m_c$. In particular, we found the position of
the zero point for the short-distance effect is in good agreement with
what we obtained via the full expression.

Nevertheless, if we still ignored the $K^*$ mass in
the region near the endpoint, the phase space would approach
$q^2=M^2$ rather than $(M-m)^2$. While it may make a slight difference
in the inclusive process (where $m$ is replaced by $m_s$),
this small change in the upper boundary of phase space
could lead to a sizeable overestimate of observable quantities
such as the decay rate and asymmetry in exclusive channels.
This is because the pole in the form factors dominates
the distribution at the endpoint.  The shift of the pole mass to $M$
would further increase the estimate.
Thus when the whole range of $q^2$ in the decay $B\rightarrow K^*$
is considered,
we apply the full expression of Eq.\ (19)
(using the experimental values for the masses $m$ and $M'$).
In this way, the integrated asymmetry for $B\rightarrow K^* l^+ l^-$
is plotted in Fig.\ 4 for monopole and dipole form factors.
We have checked that there is a variation of the zero point
of short-distance effects due to the choice of form factors.
With the same $m_t$ and $\eta$ as before, there are $s_0=0.338\mp 0.007$
(the top quark effect alone) and $\bar s_0=0.511\mp 0.017$
(the short-distance effect as a whole) for the monopole scheme, as well as
$0.291\mp 0.006$ and $0.444\mp 0.013$ for the dipole one.
The contribution of $J/\psi$ along with the $c\bar c$ effect
still dominates its resonant region.
By comparison, we find the two schemes of form factors give
similar results of the integrated asymmetry above the $\psi'$.
This is expected when both schemes use the
data of semileptonic $D$ decays as input. But asymmetries, integrated
over the interval between the $J/\psi$ and $\psi'$
and the region below the $J/\psi$ with the monopole form,
are $1.7$ times as large as that with the dipole one.
This factor becomes even larger when $q^2$ is near to the origin.
For the monopole form, a fraction of $25\%$ or so of the asymmetry occurs
in the ground state and this is far from saturating the
inclusive process. So it is anticipated that higher $K$-resonances
may play an important role in measuring asymmetry as well.

\section{Discussion and Conclusion}
As we have seen in Fig.\ 1, the magnetic-moment operator
becomes important for photons approaching the mass-shell.
The positive asymmetry arising from it is the consequence
of the coefficient $F_2^t$ being negative relative to $A-B$
in the standard model.
Thus measurement in this region will tell us the sign of $F_2^t$,
as well as confirming its bound given by the decay $b\rightarrow s\gamma$.

On the other hand, the relative phase between $F_2^t$ and $A_t+B_t$
impacts dramatically on the $b\rightarrow s l^+l^-$.
For example, a potential magnetic-moment operator
of a similar size but opposite sign to that in the standard model,
such as those allowed in supersymmetry models \cite{SUSY},
may increase the forward-backward asymmetry
of the whole kinematical region by a factor of $4.0\sim 5.0$
and a factor 2.0 for that in the low $q$ region with a cut-off
at $M_{J/\psi}$, but reversing the sign of the asymmetry.
In fact, if this relative phase is negative it shows us
evidence of new physics, whereas if it is positive it leads to
the sensitivity
of the asymmetry with respect to the $J/\psi$ and the $c\bar c$ continuum
in the middle region. We emphasize that the relative phase between
$F_2^t$ and $A_t+B_t$ manifests itself
as well in the invariant mass spectrum of dileptons.
When left-handed leptons do not mix up with right-handed ones
in the massless lepton limit,
the only interference occurs between quarks outgoing
from the four-fermion vertex and that from the penguin,
yielding a term proportional to
${\rm Re}[(A_t+B_t)F^t_2]$ in the dilepton spectrum.
As it is likely that the first experiment will be carried out
for this spectrum, we will be able to see whether there is
cancellation between the top quark components of the asymmetry
in a model independent way.

Toward high $q^2$, coefficients $A_t$ and $B_t$
become important as they are enhanced by an extra $q^2$.
The contributions of $Z$ boson penguin
along with $W$ box diagrams in this region
surpass that of the photonic penguin diagram
for a large top quark mass such as $174$~GeV (see Eq.\ (6)).
While $A-B$ (and $F_2^t$) may be determined
by the method described before, this asymmetry will help to extract
the value of the combination $A_t+B_t$ experimentally.
When $B_t$ always occurs in combinations
with the much larger $A_t$, any appreciable difference
between $A_t+B_t$ and $A_t-B_t$ is beyond the standard model.

As for the exclusive decay $B\rightarrow K^* l^+l^-$,
the difference between the two assumptions about form factors becomes
evident for low $q$.  When applied to the rare radiative decay
$B\rightarrow K^* \gamma$, the monopole form factors give
a theoretical value of $17\%$ for the ratio of the exclusive
to inclusive rate, defined as
$R=
 \displaystyle\frac{\Gamma(B\rightarrow K^*\gamma)}
			    {\Gamma(B\rightarrow X_s\gamma)}.
$
This is favored by the preliminary CLEO data\cite{xsphoton},
while the dipole scheme result of $5.7\%$ is not.
Although these two sets of form factors are similar to each other
near the endpoint, we tend to favor the prediction for
charge asymmetries obtained by using the monopole scheme,
as far as the whole kinematical region is concerned.

Finally we want to point out that possible experimental
cuts in the invariant mass of lepton-antilepton pairs
do not necessarily mean loss of statistics
as far as asymmetry is concerned.
For instance, a cut-off at the $J/\psi$ mass from which we integrate
to the endpoint $q_{\rm max}$ leads to an asymmetry larger in magnitude
than that over the whole kinematical region. In fact, this interval
gives the maxinum integrated asymmtry because of the resonant
enhancement. Our prediction for it, using $m_t=180$~GeV,
is $-1.54\times 10^{-5} \Gamma_{\rm s.l.}$ in the inclusive process,
which is accessible in the upcoming $B$ factories.

In conclusion we have presented a comprehensive analysis
of forward-backward asymmetry of $l^+$ production in
the decay $B\rightarrow X_s l^+l^-$ and $B\rightarrow K^* l^+l^-$.
We have attempted to specify certain regions in phase space of
the $B$ decay in which the asymmetry is sensitive to individual
short-distance coefficients. In particular, we note the
potential importance of the $J/\psi$, since the integrated
asymmetry near its resonant region is largely independent of
the QCD renormalization.
Given the strong dependence of
$A-B$ on the top quark mass, measurements of this asymmetry
will not only provide an effective test of the standard model
but also lead to information about $m_t$. As well,
we believe that the shape of the plots
in Figs. 1 to 4 represent well the expectation for the
forward-backward asymmetry in the framework of the standard
model. Therefore if future experiments observe any dramatic
deviation from them, they can be regarded
as the evidence of new physics beyond the standard model.

\acknowledgements

The author thanks R. Delbourgo for many discussions and encouragement
along with N. Jones for his help in preparation of this letter.
He is also grateful to the Australian Research Council for their
financial support, under grant number A69231484.

\appendix{}

\section*{Integral Formulae}

When the effect of the one-loop matrix element is studied in the main text,
we have used the notation $I_k(z)$ for the integral defined as
$$\displaystyle
I_k(z)={\rm Re}\left[\int^z_{z_0}\frac{du}{(1-u^2)^k}
					  \left(\ln\frac{1+u}{1-u}-i\pi\right)\right]
\quad\quad (k=1,2, \cdots).
$$
It holds in both ranges above and below the threshold $2m_q$.
In the latter case where $u$ is imaginary, we apply
the analytic continuation
$$\displaystyle
\arctan\left(\frac{1}{\mid u\mid}\right)
=\frac{i}{2}\left(\ln\frac{1+u}{1-u}-i\pi \right).
$$
The lower limit $z_0$ corresponds to $q^2=0$.
It is not difficult to work out $I_k$ for $k=1$
$$\displaystyle
I_1(z)=\frac{1}{2}{\rm Re}\left[\ln\frac{1+u}{1-u}
	  \left( \frac{1}{2}\ln\frac{1+u}{1-u}-i\pi\right)\right]^z_{z_0}.
$$
We would like to present in this appendix a general procedure to carry out
the integral for arbitrary $k$.
To this end, it is convenient to consider an auxiliary function
$$\displaystyle
{\cal I}(z;a)={\rm Re}\left[\int^z_{z_0}\frac{du}{(a-u^2)}
					  \left(\ln\frac{1+u}{1-u}-i\pi\right)\right]
\quad\quad (a>0),
$$
and thus the integral in question may be rewritten  as
$$\displaystyle
I_k(z)=\frac{(-1)^{(k-1)}}{(k-1)!}
	  \frac{\partial^{(k-1)}
		  {\cal I}(z;a)}
		  {\partial a^{(k-1)}}\left|_{a=1}\right. .
$$
Hence it is straightforward to derive the $I_k(z)$ for any $k$
larger than one with the explicit form of ${\cal I}(z;a)$
in terms of a series
$$\begin{array}{rl}\displaystyle
{\cal I}(z;a)
= \frac{1}{\sqrt{a}}I_1(z)
 -\frac{1}{2\sqrt{a}}\sum^\infty_{l=1} \frac{(1-\sqrt{a}\:)^l}{l}
  {\rm Re}\left\{ \right.&\displaystyle
			  J_l^{-} - J_l^{+}
			 +\frac{1}{l}\left[ \frac{1}{(1+u)^l}+\frac{1}{(1-u)^l}\right]
   \\ + & \displaystyle \left.
  \left[ \frac{1}{(1+u)^l}-\frac{1}{(1-u)^l}\right]
  (\ln\frac{1+u}{1-u}-i\pi)
  \right\}^z_{z_0},
\end{array}$$
in which $J_l^\pm $ is given by the recurrence
$$\displaystyle
J_l^\pm =\int\frac{du}{(1\pm u)^l} \frac{1}{(1\mp u)}
=\frac{1}{2}\left[\mp\frac{1}{(l-1)}\frac{1}{(1\pm u)^{l-1}}
			  +J_{l-1}^\pm\right],
$$
where we know the starting values
$$\displaystyle
J_1^+ =J_1^-=\frac{1}{2}\ln\frac{1+u}{1-u}.
$$

\newpage
{\large Figure captions}:

{\bf Figure 1}  The top quark components of forward-backward charge
			  asymmetry in the decay $B\rightarrow X_s\; l^+l^-$,
			  {\em integrated} over the region of $[0,\; q^2]$.
			 The dashed line corresponds to the photon penguin
			  diagram, while the solid to $Z$ penguin plus
			  $W$ box diagrams.
			 We take the values $m_t=174$~GeV, $\eta=1.75$,
			  and $m_b=4.9$~GeV and the limit of
			  $m_s^2/m_b^2\rightarrow 0$.
			 Both plots are
			  rescaled by the semileptonic decay width of the $B$ meson,
			  multiplied by the factor of $10^5$.

{\bf Figure 2}  Integrated asymmetries due to short-distance effects
			  in $B\rightarrow X_s\; l^+l^-$.
			 The dashed line corresponds to the active
			  $q\bar q$ contribution dominated by the $c\bar c$,
			  and the solid to the top
			  quark one, that is the sum of plots in Fig.\ 1.
			 We use $m_c=1.5$~GeV and the remaining
			  parameters and the scale are the same as in Fig.\ 1.

{\bf Figure 3} The same as in Figs.\ 1 and 2 for the sum (dashed line)
			 of all effects (see Eq.\ (10) in the main text),
			 including resonant contributions (solid line).

{\bf Figure 4}  Integrated forward-backward asymmetries
			  in the decay $B\rightarrow K^*\; l^+l^-$.
			 The dashed line is associated with the monopole
			  scheme of form factors, while the solid
			  line goes with the dipole one.
			 We take $M=5.28$~GeV, $m=0.892$~GeV, and
			  $M'=5.38$~GeV.
			 The remaining
			  parameters and the scale are the same as in previous
			  figures.
 \newpage
\setlength{\unitlength}{1cm}

\begin{figure}
\begin{picture}(16,20)
\put (0.5,2.0){\epsfxsize=13cm \epsfbox{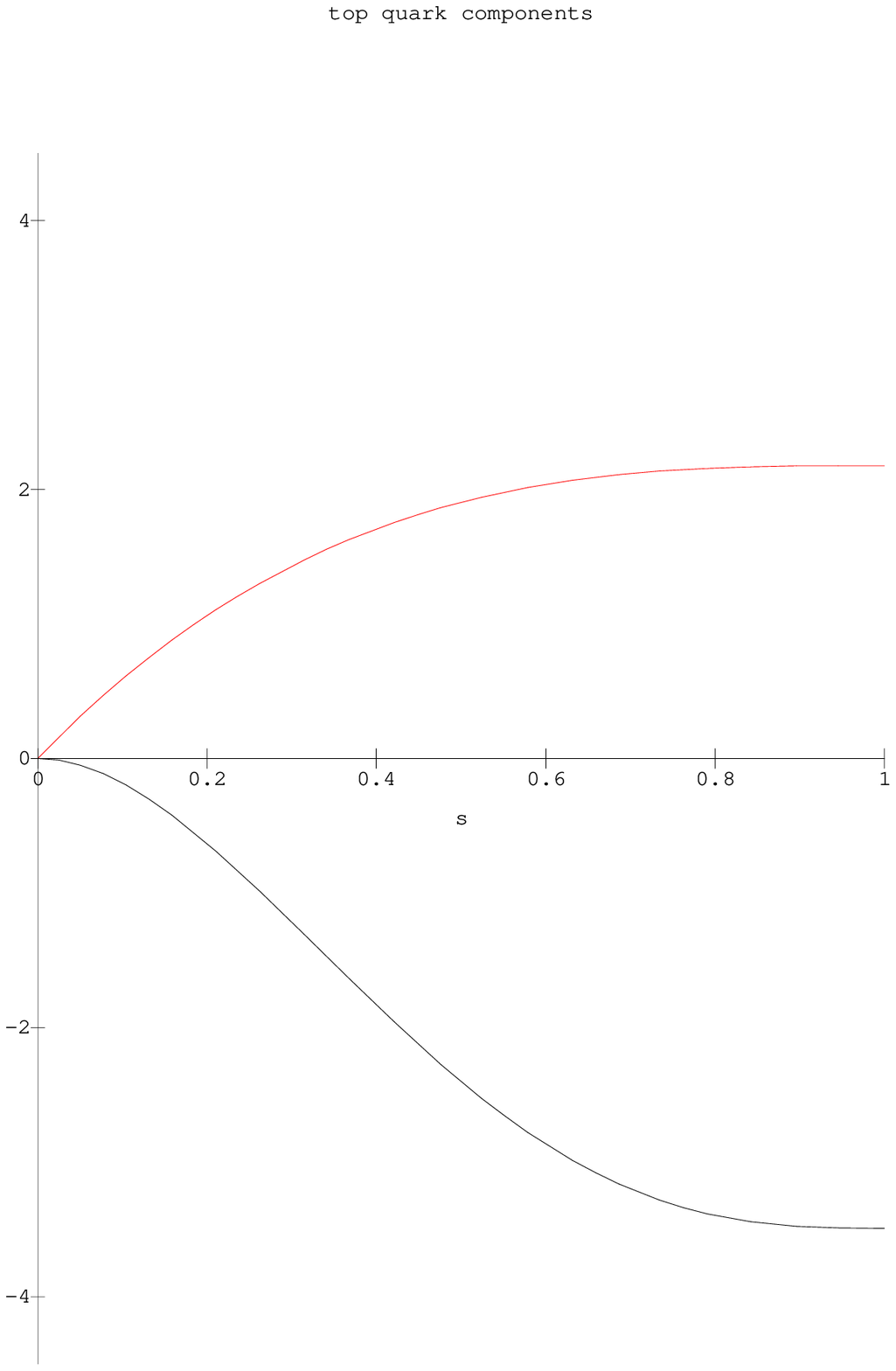}}
\put (0.5,18.0){$\displaystyle\frac{\Gamma_{\rm FB}}{\Gamma_{\rm s.l.}}$}
\end{picture}
\caption{D. Liu}
\label{fig:f1}
\end{figure}

\newpage
\begin{figure}
\begin{picture}(16,20)
\put (0.5,2.0){\epsfxsize=13cm \epsfbox{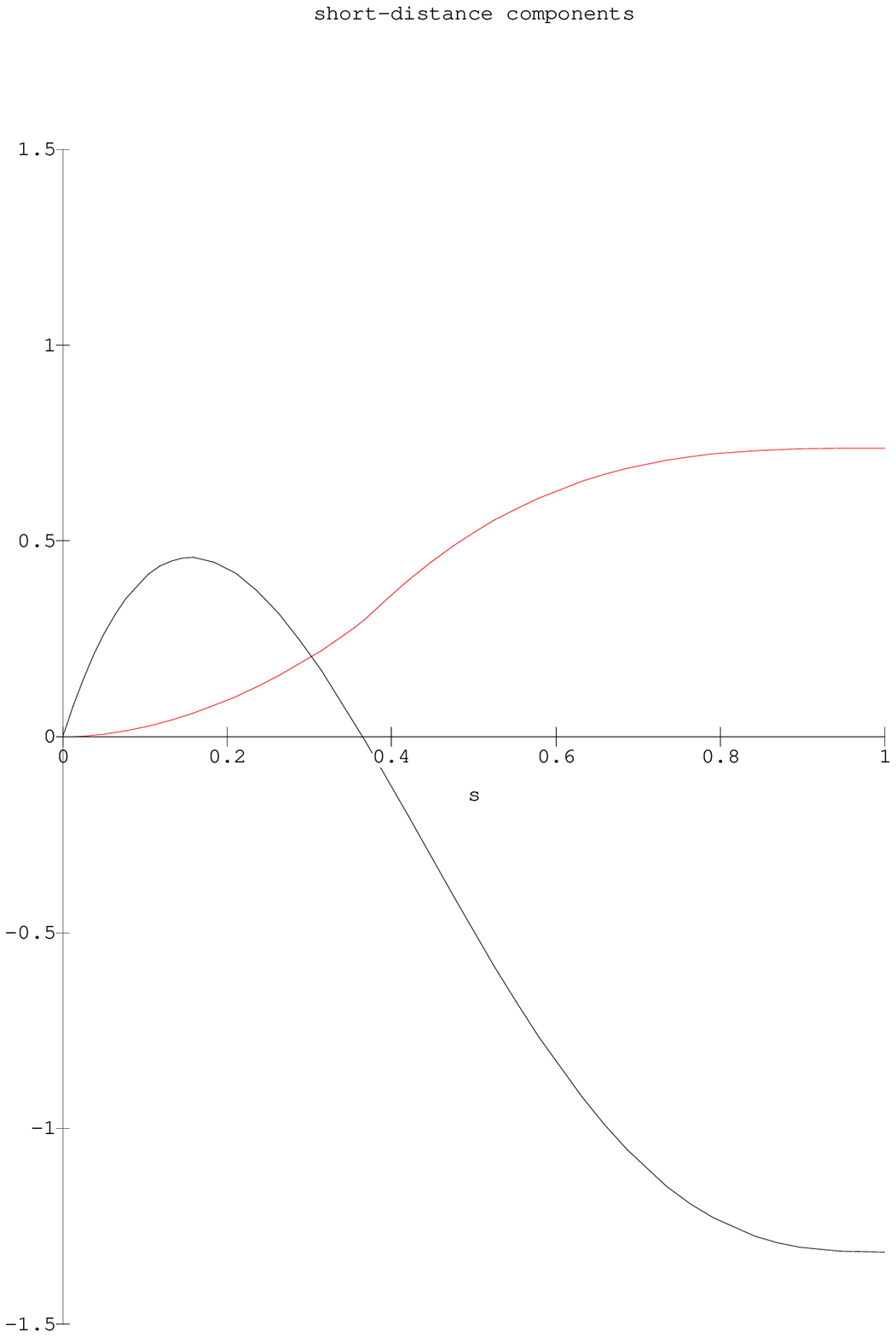}}
\put (0.5,18.0){$\displaystyle\frac{\Gamma_{\rm FB}}{\Gamma_{\rm s.l.}}$}
\end{picture}
\caption{D. Liu}
\label{fig:f2}
\end{figure}

\newpage
\begin{figure}
\begin{picture}(16,20)
\put (0.5,2.0){\epsfxsize=13cm \epsfbox{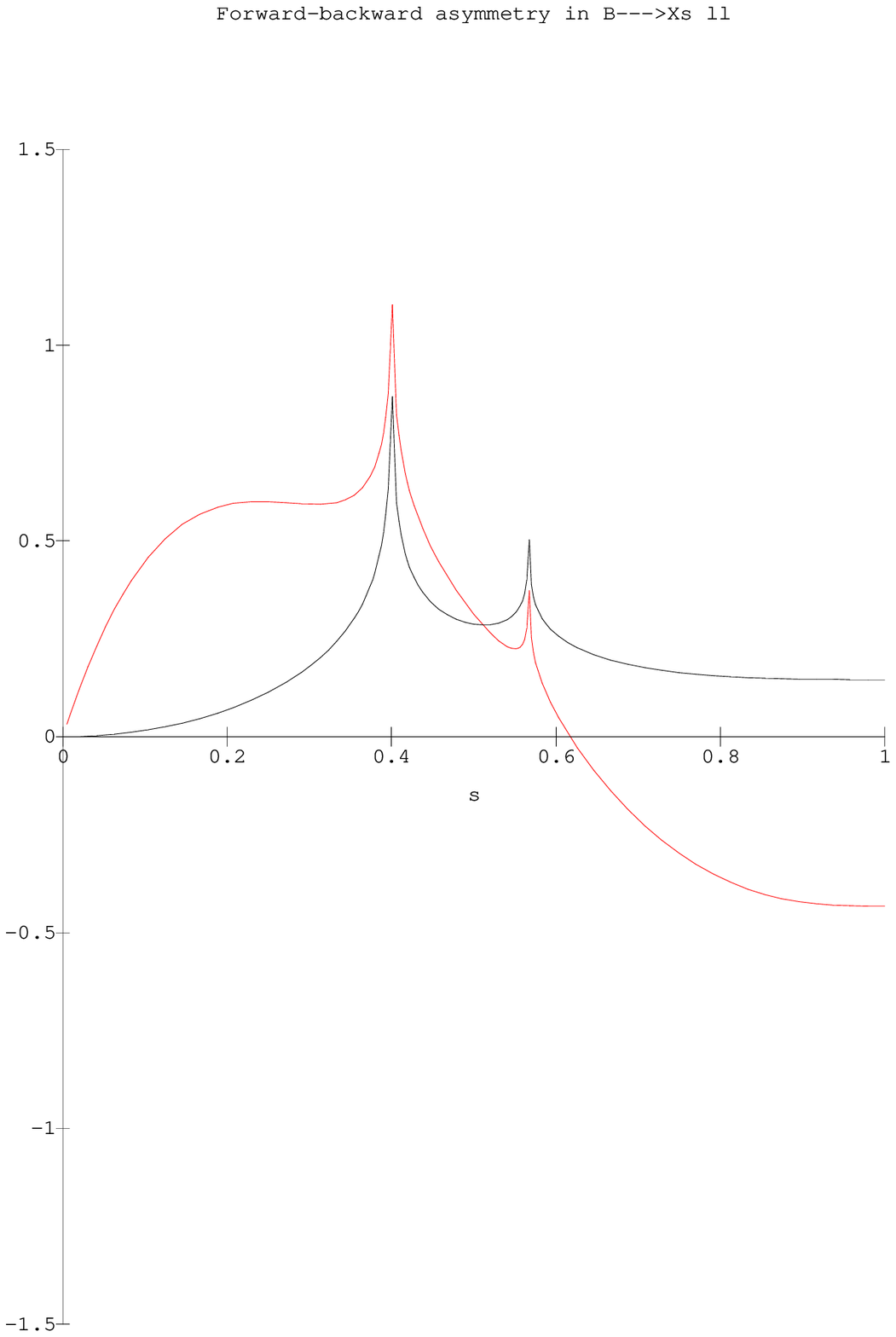}}
\put (0.5,18.0){$\displaystyle\frac{\Gamma_{\rm FB}}{\Gamma_{\rm s.l.}}$}
\end{picture}
\caption{D. Liu}
\label{fig:f3}
\end{figure}

\newpage
\begin{figure}
\begin{picture}(16,20)
\put (0.5,2.0){\epsfxsize=13cm \epsfbox{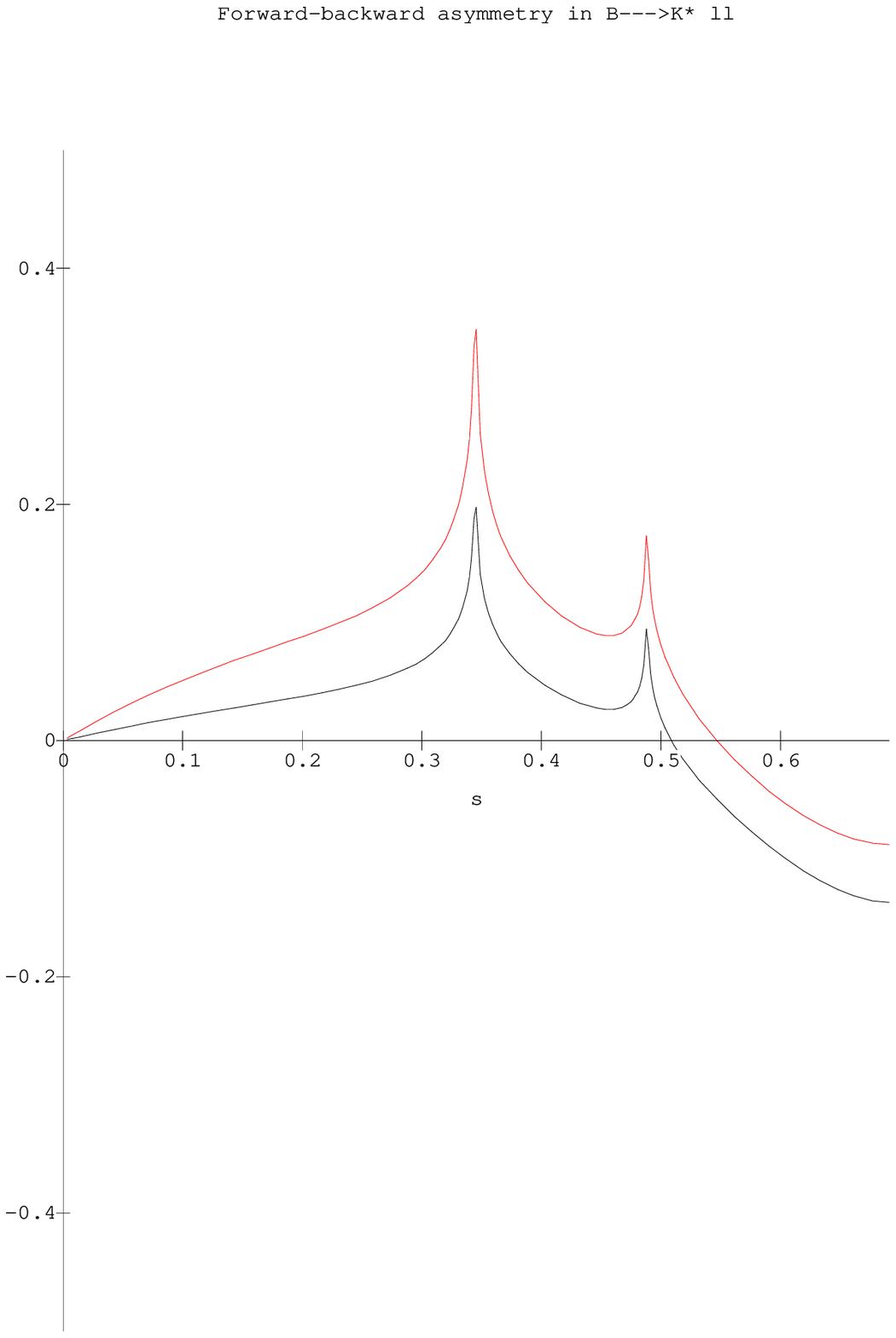}}
\put (0.5,18.0){$\displaystyle\frac{\Gamma_{\rm FB}}{\Gamma_{\rm s.l.}}$}
\end{picture}
\caption{D. Liu}
\label{fig:f4}
\end{figure}


\begin{references}
\bibitem{NLO} The anomalous dimesions $a_i$ along with the coefficients
		    $p_i$, $s_i$, and $r_i$ as well as afterwards,
		    $h_i$ and $c_i$ are given in refs.~\cite{misiak,munchen}.
\end{references}
\end{document}